# Dynamics of Non-Autonomous Oscillator with a Controlled Phase and Frequency of External Forcing

D.A. Krylosova[2], E.P. Seleznev[1,2], N.V. Stankevich[1,3,4]

[1] *Kotel'nikov's Institute of Radio-Engineering and Electronics of RAS, Saratov Branch, Zelenaya 38, Saratov, 410019, Russia*
[2] *Saratov State University, Astrakhanskaya, 83, 410012, Russia*
[3] *Yuri Gagarin State Technical University of Saratov, Politehnicheskaya 77, Saratov, 410054, Russia*
[4] *St.Petersburg State University, Universitetskiy proezd, 28, Peterhof, 198504, Russia*

The dynamics of a non-autonomous oscillator in which the phase and frequency of the external force depend on the dynamical variable is studied. Such a control of the phase and frequency of the external force leads to the appearance of complex chaotic dynamics in the behavior of oscillator. A hierarchy of various periodic and chaotic oscillations is observed. The paper studies the structure of the space of control parameters. It is shown that in the dynamics of the system there are oscillatory modes similar to those of a non-autonomous oscillator with a potential in the form of a periodic function, but there are also significant differences.

Keywords: non-autonomous oscillator, sinusoidal-nonlinearity, chaos

**Introduction.**

Many systems, including radiophysical, biological, and others, exhibit oscillatory processes in which one the system acts on another with a periodic signal, but when the operating conditions change, the frequency of forcing changes. For example, in information transmission systems to ensure high stability, the so-called phase-locked loop is used [1-3]. The system of cardiovascular regulation of living organisms with a change in load increases or decreases the heart rate [4-8]. In such interactions, the dependence of the phase or frequency on the dynamic variable can lead to the appearance of complex dynamics in the system. The control process in this case is very complex, its research and modeling encounters a number of difficulties. One of the ways in the study of such systems and processes is the consideration of simpler objects in which the excitation of oscillations and frequency control are rather easily modeled. As such a system, it is convenient to use the classical model of the theory of oscillations - a linear oscillator under external harmonic force.

In the framework of this work, a study of the dynamics of a non-autonomous oscillator with a controlled phase and frequency of external force is presented. The structure of the space of control parameters is investigated. The role of parameters is determined. The work is structured as follows. Section 2 describes the object of study: a linear oscillator with an external periodic force, the external force of which has a frequency and phase dependent on a dynamical variable, which leads to the appearance of nonlinearity in the system. Section 3 discusses in detail the dynamics of the oscillator, the phase of which depends on the dynamical variable. Section 4 presents the study of an oscillator with a frequency depending on a dynamical variable.

**2. Object of study: a linear oscillator under external periodic influence and an oscillator with a controlled phase and frequency.**

As the simplest object of study, we will choose the classical model of the theory of oscillations [9, 10] – *RLC* - circuit, excited by an external signal, which is written in the following form:

$$\ddot{x} + 2\alpha\dot{x} + \omega_0^2 x = A Sin(\omega t + \varphi), \tag{1}$$

where $x, \dot{x}$ are the dynamic variables, α is the dissipation coefficient, $\omega_0$ is the natural frequency of the oscillations of the circuit, *A*, $\omega$ and $\varphi$ are the amplitude, frequency and phase of the external forcing. Equation (1) describes the behavior of a linear non-autonomous oscillator, the dynamics of which well known: the external force in such a system excites periodic self-oscillations [9, 10].

The complication of the dynamics of such a system is traditionally observed with the addition of nonlinear terms. For example, adding cubic nonlinearity will result in the well-known Duffing oscillator [11-15]. By adding exponential nonlinearity, we get a Toda oscillator [16-17], whose equation reproduces the dynamics of the *RL*-diode circuit [18-21]. When such systems interact, they exhibit complex dynamic behavior, including chaos, quasiperiodic oscillations, multistability, nonlinear resonance, etc. [22-30].

In the framework of this work, we consider the features of the system when the external signal is complicated, i.e. taking into account the dependence of the phase and frequency of external influence on the dynamic variable.

The first version of the model corresponds to the case when the phase depends on the variable, as result the equation (1) becomes non-linear. The dependence of the phase on the variable is the simplest one, i.e. through a linear function:

$$\varphi(x) = kx, \tag{2}$$

where *k* is a constant coefficient, *x*(*t*) is a dynamic variable.

After the transition to dimensionless time $\tau = \omega_0 t$, equation (1) takes the form

$$\ddot{x} + 2\alpha\dot{x} + x = A Sin(p\tau + kx), \tag{3}$$

where $p = \omega/\omega_0$ is the normalized frequency of external force. Equation (3) contains a nonlinearity of type Sin*(kx)*.

The second version of the model that we will consider in the framework of this work is an oscillator, the frequency of which depends on the dynamical variable. The equation of forced oscillations of the oscillator in this case has the form

$$\ddot{x} + 2\alpha\dot{x} + x = A Sin(p(x)\tau + \varphi), \tag{4}$$

where *x* is the dynamic variable, α is the dissipation coefficient, $\omega_0$ is the eigenfrequency of the oscillator, *A* is the amplitude, *p* is the frequency, and $\varphi$ is the phase of the external force, respectively. As in the previous case, we assume that the dependence of the frequency of the external force on the dynamic variable is linear:

$$p(x) = p_0 + kx(t). \qquad (5)$$

Then equation (1) takes the form:
$$\ddot{x} + 2\alpha\dot{x} + x = A\mathrm{Sin}((p_0 + kx)\tau + \varphi). \qquad (6)$$

Setting $\varphi = 0$, we obtain an equation of the form
$$\ddot{x} + 2\alpha\dot{x} + x = A\mathrm{Sin}((p_0 + kx)\tau). \qquad (7)$$

Thus, by controlling the phase and frequency of the external force, the linear equation describing the forced oscillations of the linear oscillator is also converted into a nonlinear one with nonlinearity of the $\mathrm{Sin}(kx)$ type.

An oscillator with nonlinearity of the sine type is one of the reference models of nonlinear dynamics [10]. It describes the oscillations of a mathematical and physical pendulum, and also appears in other applied problems, for example, when considering Josephson contact [31-34], when studying self-induced transparency in nonlinear optics, when analyzing the bending of an elastic beam, etc. Dynamical systems of this type have very rich dynamics [35-39]. The richness of the dynamics of such oscillators is related to the form of the potential function, which is periodic and has an infinite number of maxima and minima. In the case of a symmetric potential function, dynamics is observed, accompanied by a transition from one well to another. In addition to the well-known scenarios of transition to chaos and types of bifurcations, so-called metastable chaos takes place in such systems. In the case of asymmetry of the potential function, the so-called particle drift in the periodic potential is observed.

The aim of this work is to numerically study the forced oscillations of a linear oscillator when controlling the phase and frequency of the driving force using equations (3) and (7) as an example.

### 3. Numerical analysis of the dynamics of a controlled phase system.

The analysis of the nature of the forced oscillations in the work was carried out on the basis of an assessment of the spectrum of Lyapunov exponents, as well as on the analysis of phase portraits in the stroboscopic section. The amplitude A, frequency p, and phase control coefficient k were used as control parameters. For stability analysis, equation (3) was transformed into a system of three first-order differential equations:
$$\begin{aligned}\dot{x} &= y, \\ \dot{y} &= -2\alpha y - x + A\mathrm{Sin}(z), \\ \dot{z} &= p + ky \end{aligned} \qquad (8)$$

We study the characteristic structure of various parameter planes in this case.

In fig. Figure 1 presents charts of the dynamic regimes of system (8) on the plane of parameters $(k, A)$ for three different values of the frequency of forcing $p$. Different colors denote the regions of periodic regimes with different periods and chaotic oscillations; under the figure, the corresponding color palette is presented. Fig. 1a illustrates the structure of the parameter plane $(k, A)$ at $p = 0.25$. In the dynamics of system (8), a sequence of period doubling bifurcations is observed, ending with a transition to chaos. In the region of the existence of chaos, its development is observed, associated with a decrease in the connectivity of the chaotic attractor, alternating with the appearance of zones of periodic oscillations. Fig. 1b illustrates the structure of the parameter plane ($k$, $A$) at $p = 1$. The structure of the parameter plane remains qualitatively unchanged, only the bifurcation values of the parameters A and k change. Fig. 1c illustrates the structure of the parameter plane ($k$, $A$) at $p = 5$. In general, the plane structure also represents an alternation of zones of periodic and chaotic oscillations. However, bands of periodic regimes corresponding to higher resonances appear. As can be seen from Fig. 1, an increase in the parameters $A$ and $k$ leads to a qualitatively identical change in the dynamics of the system.

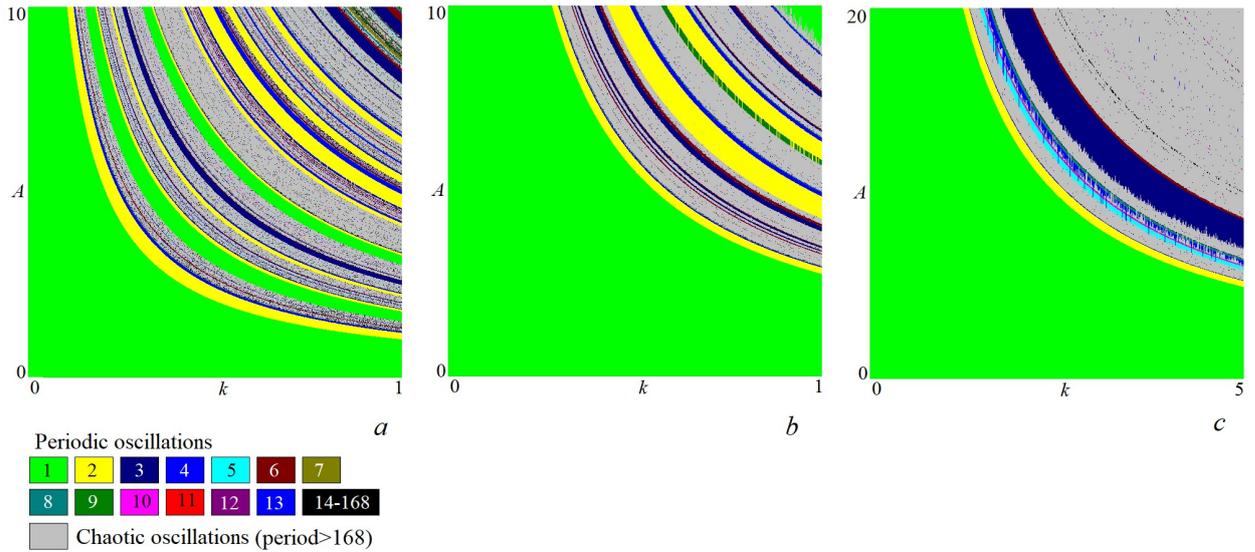

Fig. 1. Charts of dynamical modes of model (8) for α = 0.1, a) $p = 0.25$, b) $p = 1$, c) $p = 5$.

Figure 2 shows phase portraits illustrating the appearance of a chaotic attractor as a result of a cascade of period doubling bifurcations. Figure 2a shows the limit cycle for small values of the parameters of the amplitude of the external signal and coefficient $k$. With an increase in the amplitude of the external force, the limit cycle increases in size (Fig. 2b). As the coefficient $k$ increases, the shape of the limit cycle changes, additional loops appear (Fig. 2c), however, in the stroboscopic section, this attractor still corresponds to a single fixed point. Figure 1d shows an example of a more complex attractor for large values of the parameters $A$ and $k$. On the basis of such a limit cycle, a cascade of period doubling bifurcations occurs in the system. In Fig. 2e, an example of a double limit cycle is shown, and then the chaos develops, presented in Fig. 2e. Inside the chaos region with a further increase in the parameters on the parameter plane, periodicity windows are observed, inside which a cascade of period doubling bifurcations also occurs and chaos also occurs. For all cases shown in Fig. 2, the dynamics of the system develops in the vicinity of one of the potential wells located on one side of the unstable zero equilibrium state. In this case, the phase trajectory can also enter the region of the second symmetric potential well, but then returns. This is clearly seen in the stroboscopic sections, which are located in the negative region of the dynamical variable $x$.

A similar scenario is observed with increasing frequency parameter $p$. However, with increasing frequency, it grows in size and begins to visit other potential wells that are more distant from the equilibrium state. At the same time, dynamic chaos developing at various frequencies has its own characteristic features. To analyze the features, stroboscopic sections of phase portraits and Fourier spectra for chaotic attractors were constructed for various values of the parameter $p$, which are presented in Fig. 3.

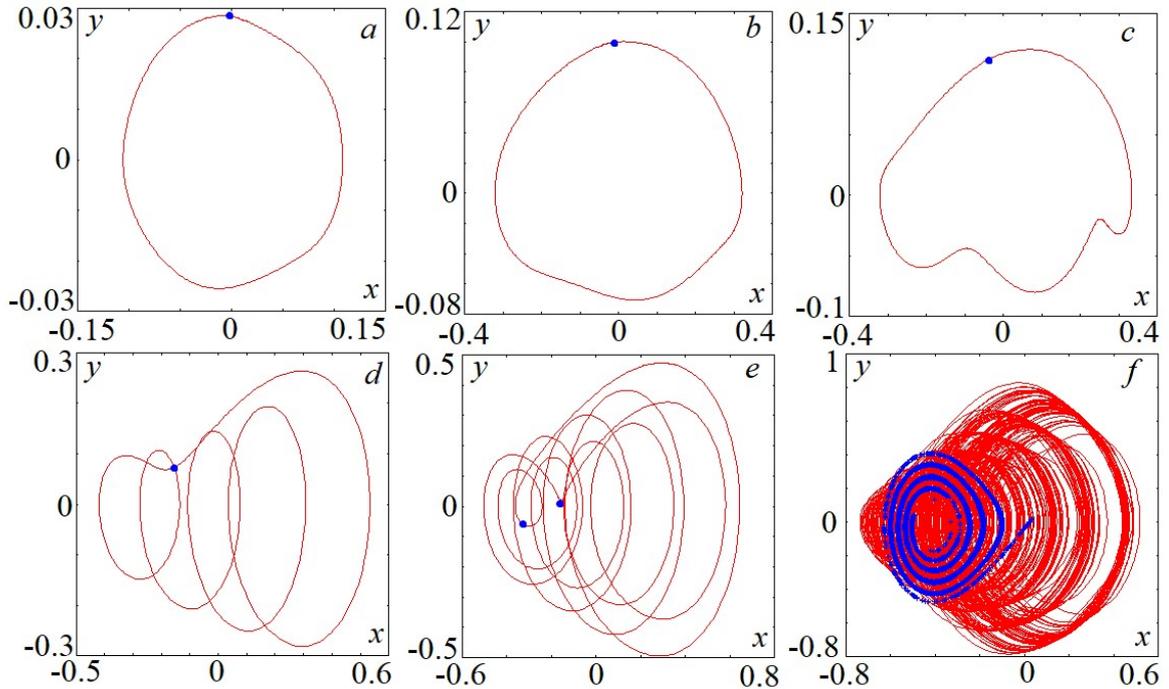

Fig. 2. Two-dimensional projections of phase portraits (red lines) and cross-sectional stroboscopic sections (blue dots) for various points of the parameter plane ($k$, $A$) at $p = 0.25$. a) $k = 0.1$, $A = 0.5$; b) $k = 0.3$, $A = 0.5$; c) $k = 0.3$, $A = 0.75$; d) $k = 0.32$, $A = 2.67$; e) $k = 0.36$, $A = 3.07$; f) $k = 0.62$, $A = 5.85$.

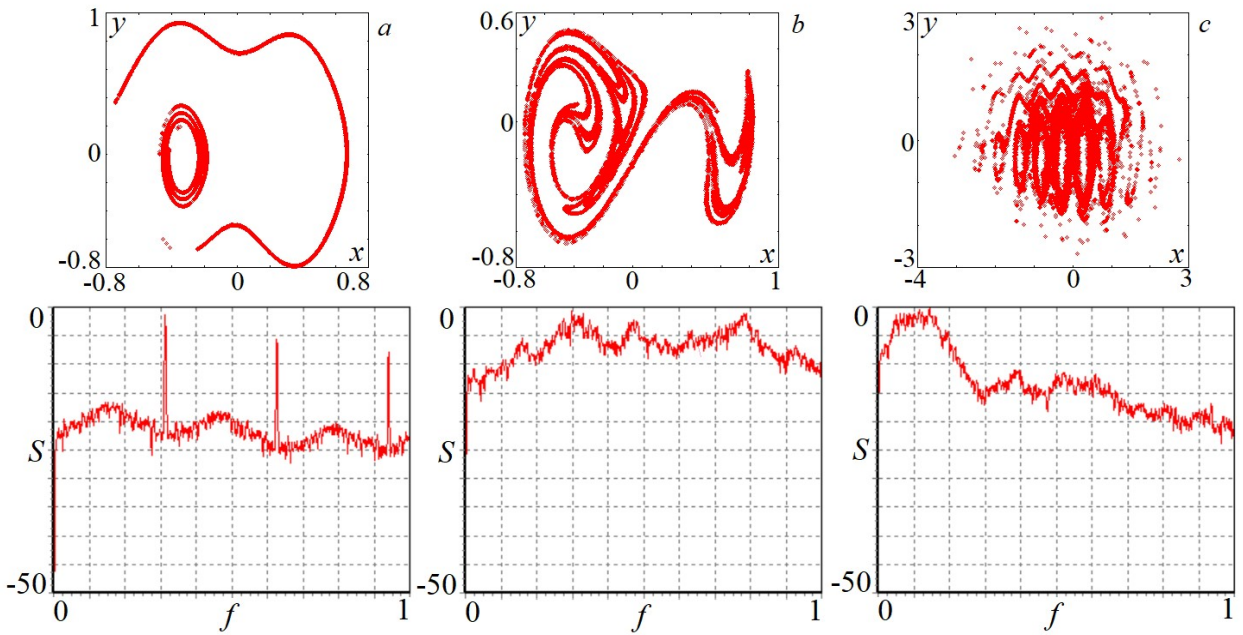

Fig. 3. Stroboscopic cross sections and Fourier spectra for various chaotic modes. a) $p = 0.25$, $k = 0.85$, $A = 8.18$; b) $p = 1$, $k = 0.57$, $A = 5.35$; c) $p = 5$, $k = 3.03$, $A = 11.1$.

For small values of the parameter p, which is responsible for the frequency of the external force ($p = 0.25$), the oscillator dynamics mainly develops in one of the potential wells close to the zero equilibrium point. The peak corresponding to the base frequency of the limit cycle from which this chaotic attractor was born is clearly visible in the Fourier spectrum. With an increase in the frequency of external force *p* ($p = 1$), the attractor becomes more developed and jumps are observed in dynamics from one potential well to another. The Fourier spectrum of such a regime is broadband and does not contain individual peaks, as it was for the previous case. At $p = 5$, the phase portrait looks even more developed; the phase trajectory visits about 7 potential wells. The Fourier spectrum is also broadband, but in this case a certain higher-amplitude band appears at low frequencies, which corresponds to filtering the signal by the circuit at the resonant frequency.

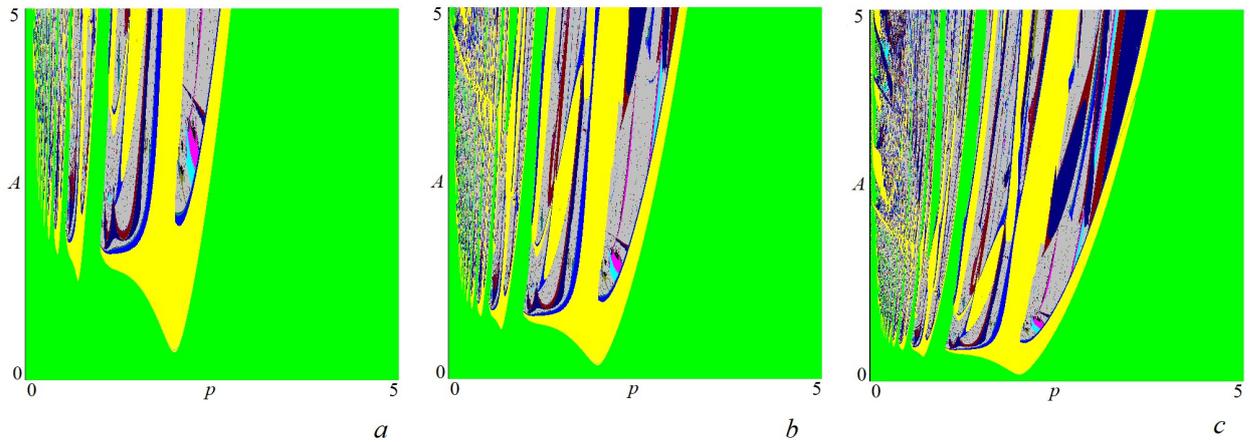

Fig. 4. Charts of dynamical modes of model (8) with α = 0.1, a) k = 0.5, b) k = 1, c) k = 2.

Now we turn to the study of the characteristics of the plane of parameters corresponding to the most classical in terms of synchronization: frequency - amplitude of the external signal. As a frequency parameter, we will use the parameter *p*. Figure 4 presents charts of the modes of oscillations of system (8) on the plane of parameters (*p, A*) for various values of the parameter k; the color palette is similar to Fig. 1. Figure 4a illustrates the structure of the parameter plane (*p, A*) at *k* = 0.5. Here it is possible to distinguish separate zones of complex behavior associated with the so-called resonances at higher harmonics. At low frequencies on the parameter plane, in the dynamics of the system, a sequence of period doubling bifurcations is observed, ending with a transition to chaos. The lines of bifurcations of period doubling have the characteristic form of tongues with a certain threshold in the parameter *k*. So, for the first line of period doubling (at the maximum frequency of the external signal *p*), the minimum is located at the doubled resonant frequency, which is typical for the structure of the space of control parameters of a non-autonomous nonlinear oscillator [9-12]. With a decrease in the frequency of forcing, similar lines of period doubling are observed at frequencies corresponding to subresonances. As the frequency decreases, the threshold for bifurcation of the doubling period increases. With an increase in the parameter k, chaos develops due to a decrease in the connectivity of the chaotic attractor, alternating with the appearance of zones of periodic oscillations. On the whole, the structure of the plane of control parameters (Fig. 4a) is in many ways similar to that for a non-autonomous nonlinear oscillator [18, 19]: it is possible to distinguish separate zones of complex behavior associated with the so-called resonances at higher harmonics.

An increase in the parameter k (Fig. 4b and Fig. 4c) leads to an increase in the range of variation of the phase of the forcing, and as a result to an increase in the regions of complex behavior and the complication of their structure.

Figure 5 presents charts of dynamic modes on the plane of parameters (*p, k*) for two values of parameter *A*: *A* = 1 and *A* = 10. Qualitatively, the structure of the parameter plane repeats the analogous one presented in Fig. 4. Which also confirms that a change in the parameters *A* and *k* qualitatively leads to the same result. Figure 5 illustrates the diversity of the zones of existence of various modes of oscillation.

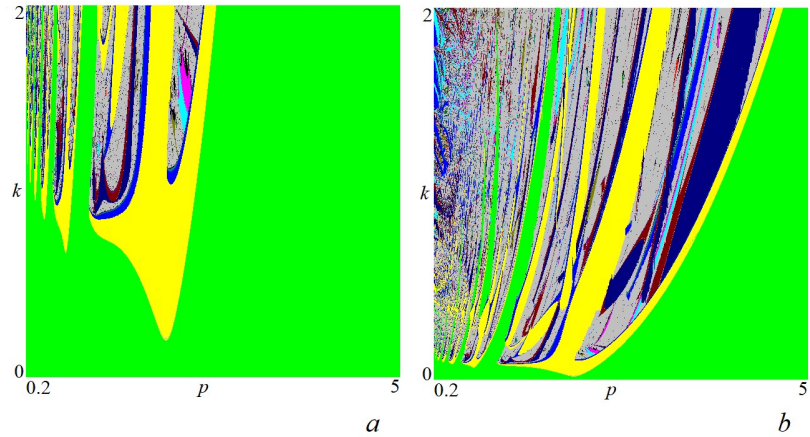

Fig. 5. Charts of dynamical modes of model (8) with α = 0.1, a) A = 1, b) A = 10.

**4. Numerical analysis of the dynamics of a system with a controlled frequency**

In the case of frequency control, the dynamics does not change qualitatively; the main difference in the system behavior is the structure of the space of control parameters. A change in the frequency of forcing, which is a consequence of its dependence on a dynamical variable, leads to the fact that the instantaneous frequency of the oscillations changes and the nature of the oscillations is more complex than when controlling the phase. The system of first-order equations in this case has the form:

$$\dot{x} = y,$$
$$\dot{y} = -2\alpha y - x + A Sin(z), \qquad (9)$$
$$\dot{z} = (p_0 + kx) + ky\tau.$$

It should also be noted here that, unlike the system of equations (8) in (9), the variable *z* clearly depends on the time *τ*.

The study of model (9) will be carried out similarly to a controlled phase system. To analyze the dynamics when varying the parameters, we also use the dynamic mode map method, which was also verified by analyzing the spectrum of Lyapunov exponents.

Figure 6 shows the charts of the modes of oscillations of system (9) on the plane of parameters (k, A) for various values of the parameter $p_0$ and the dissipation parameter *r* = 0.1. The color palette used was the same as for Fig. 1. Fig. 6a illustrates the structure of the parameter plane

($k$, $A$) at $p_0 = 0.25$. The overall picture remains the same: on the parameter plane, self-oscillation bands with a period of 1 in the stroboscopic section are observed. With basic limit cycles, a cascade of period doubling bifurcations occurs and a chaotic attractor arises. The difference from the case of the dependence of the phase on the variable is that these structures are observed only at small values of the parameters A and k. For large parameter values, the periodicity windows become very narrow and the chaotic dynamics mode dominates. Fig. 6b illustrates the structure of the parameter plane ($k$, $A$) with $p_0 = 1$. The structure of the parameter plane remains qualitatively unchanged, bands of periodic regimes are observed, from which chaotic oscillations arise through a cascade of period doubling bifurcations. However, for this choice of parameters, the regions of the limit cycle become more pronounced. With a further increase in the parameter p0, the chaos regions disappear and only periodic oscillations are realized in the system. This effect is due to the fact that with an increase in $p_0$ the amplitude of the forced oscillations decreases, and, accordingly, the amplitude of the change in the forcing frequency.

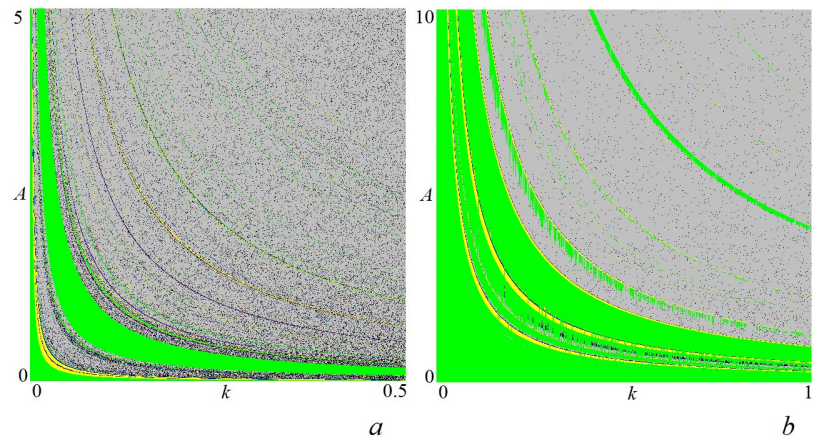

Fig. 6. Charts of the dynamic modes of model (9) for $\alpha = 0.1$, a) $p_0 = 0.25$, b) $p_0 = 1$.

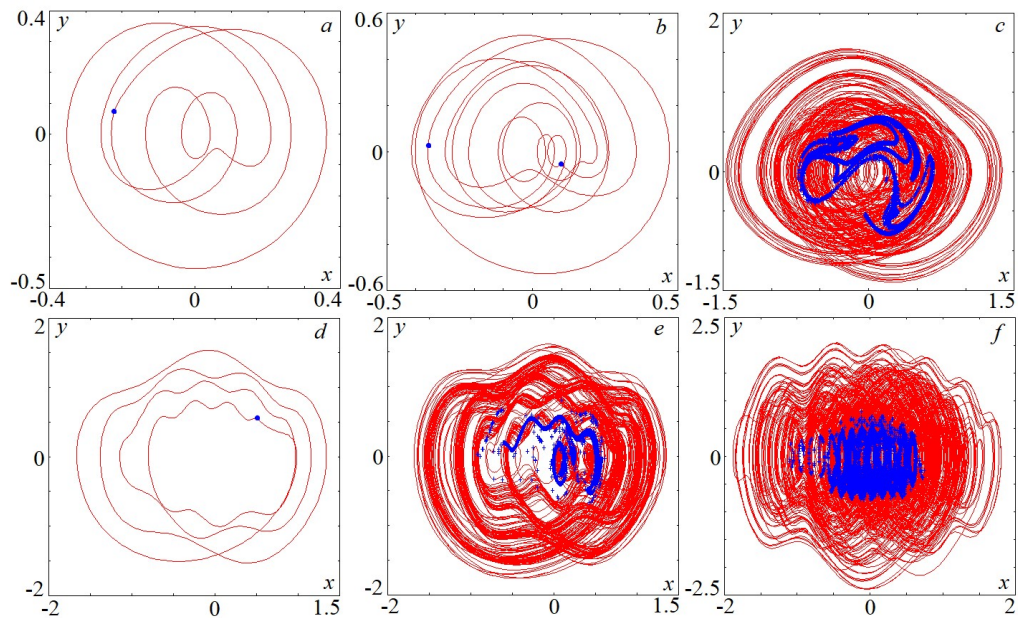

Fig. 7. Two-dimensional projections of phase portraits (red lines) and stroboscopic sections (blue dots) for various points of the parameter plane ($k$, $A$) at $p_0 = 0.25$. a) $k = 0.02$, $A = 0.167$; b) $k = 0.03$, $A = 0.2$; c) $k = 0.045$, $A = 0.417$; d) $k = 0.085$, $A = 0.733$; e) $k = 0.11$, $A = 0.9$; e) $k = 0.168$, $A = 1.35$; g) $k = 0.283$, $A = 2.267$.

Figure 7 shows examples of projections of phase portraits and their stroboscopic sections for the case $p_0 = 0.25$. The phase portrait for the limit cycle of the period is a multi-turn cycle, which corresponds to one point in the stroboscopic section (Fig. 7a). On its basis, a cascade of period-doubling bifurcations occurs. The phase portrait in Fig. 7b corresponds to a double limit cycle; in Fig. 7c, an example of a chaotic attractor is presented. For small values of the parameters $A$ and $k$, the oscillations occur inside two potential wells of the oscillator, the imaging point visits the vicinity of each of the wells. With increasing parameters $A$ and $k$, a larger number of potential wells are involved in the dynamics of the system. Figure 7d shows an example of the limit cycle for such a case. The stroboscopic sections shown in Figs. 7e and 7e clearly show the gradual involvement of more potential wells in the dynamics.

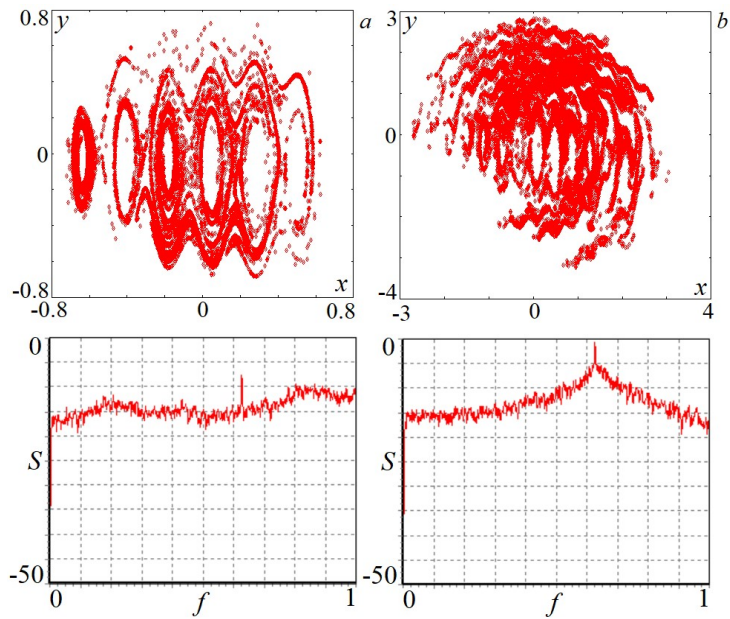

Fig. 8. Stroboscopic cross sections and Fourier spectra for various chaotic modes. a) $p = 0.25$, $k = 0.168$, $A = 1.35$; b) $p=1$, $k = 0.39$, $A = 2.77$.

An increase in the frequency parameter $p_0$ also affects the spectral characteristics of the dynamic mode. Figure 8 shows examples of phase portraits in the stroboscopic section and Fourier spectra for chaotic attractors at two different values of the frequency parameters $p_0$. It is clearly seen in phase portraits that with increasing frequency the attractor becomes more developed and the phase trajectory moves on the basis of a larger number of potential wells. For both cases, the spectrum is broadband, but there is a pronounced component corresponding to the base oscillation frequency. In the case of the regime shown in Fig. 8a, a relatively high uniformity of the spectrum should be noted. Perhaps by the selection of control parameters it can realized chaotic modes with a uniform spectrum.

Next, we consider the structure of other parameter planes for a non-autonomous oscillator, the frequency of which depends on the dynamic variable. In fig. Figure 9 presents charts of the dynamic regimes of system (9) on the plane of the frequency – amplitude parameters of the external force ($p_0$, $A$) for various values of the frequency tuning parameter k and the dissipation parameter

$r = 0.1$. Fig. 9a illustrates the structure of the parameter plane ($p_0$, $A$) for $k = 1$. In the structure of the parameter plane, there is some similarity with Fig. 4a, in the case of controlling the phase of exposure, however, there are significant differences. The region of chaotic dynamics, as well as for the model with phase adjustment, is limited by twice the resonant frequency. In the parameter plane, the structure within which a cascade of period doubling bifurcations with the formation of a chaotic attractor, which is located between the resonant and doubled resonant frequencies, is clearly expressed. The period doubling bifurcation line has a minimum near the doubled resonant frequency. At a frequency of external force of a lower resonance frequency, at small amplitudes of the impact, self-oscillations are destroyed and a chaotic attractor appears, and there is no system of cascades of period doubling corresponding to subresonances, as was the case with a phase-tuning system. With an increase in the frequency tuning parameter $k$ ($k = 1$, Fig. 9b), the threshold for the appearance of bifurcation of period doubling near the doubled frequency becomes smaller, the region of chaotic oscillations expands to the region of high frequencies of external influence. Moreover, in the region of more than twice the resonant frequency, new cascades of period doubling are formed at multiple resonant frequencies, but with a large threshold in amplitude. With a further increase in the frequency tuning parameter $k$ ($k = 2$, Fig. 9c), the cascade of period doubling bifurcations in the vicinity of the doubled resonant frequency expands to high frequencies and, when it crosses the triple frequency, a single region of chaos is formed.

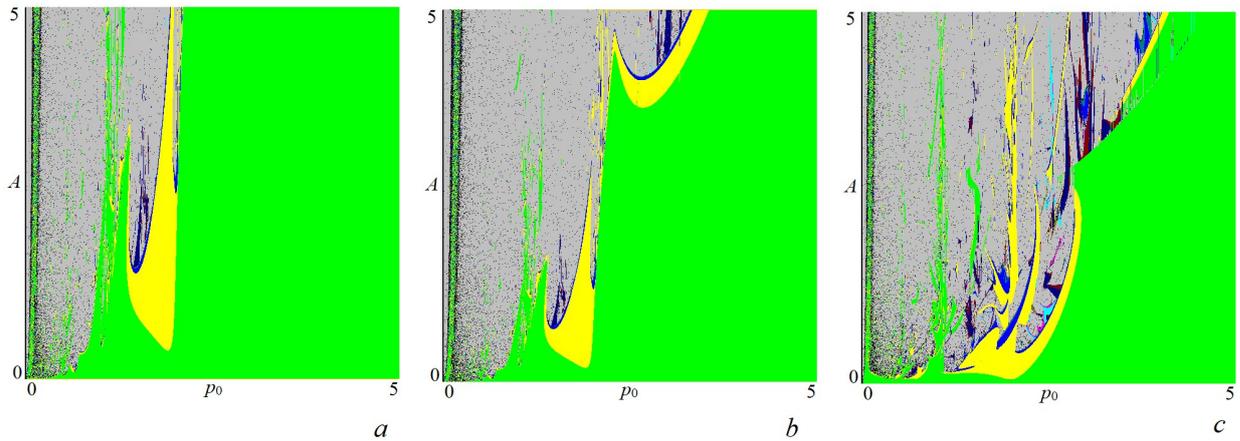

Fig. 9. Charts of dynamical modes of model (9) with $\alpha = 0.1$, a) $k = 0.5$, b) $k = 1$, c) $k = 2$.

Fig. 10 shows the charts of the dynamic modes of system (9) on the plane of parameters, the frequency of the impact is the frequency control coefficient ($p_0$, $k$) for various values of parameter $A$. Fig.10a illustrates the structure of the parameter plane ($p_0$, $k$) for $A = 1$. In the structure of the parameter plane, there is some similarity with Fig. 9a, as well as for the case of controlling the phase of the force. The region of chaotic dynamics is limited by twice the resonant frequency. On the parameter plane, the structure within which a cascade of period doubling bifurcations with the formation of a chaotic attractor, which is located between the resonant and doubled resonant frequencies, is clearly expressed. At a frequency of external force of a lower resonance frequency, at small amplitudes of the impact, self-oscillations are destroyed and a chaotic attractor appears, and there is no system of cascades of period doubling corresponding to subresonances, as was the case with a phase-tuning system. At frequencies greater than twice the resonant frequency, periodic self-oscillations are observed. With increasing parameter $A = 10$ (Fig. 10b), the structure of the parameter plane changes significantly. The period doubling bifurcation threshold near the doubled resonant frequency becomes much smaller. In the vicinity of the tripled resonant frequency, one more period doubling line is observed, with an increase in the frequency tuning parameter $k$ over a wide range of external forcing frequencies, a cascade of period doubling bifurcations is observed and chaotic dynamics appear at frequencies doubled.

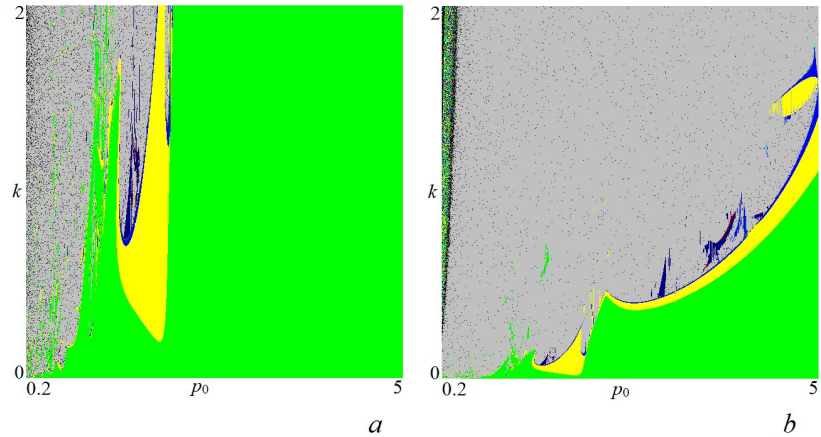

Fig. 10. Charts of dynamical modes of model (9) with $\alpha = 0.1$, a) $A = 1$, b) $A = 10$.

## 5. Conclusion.

Thus, the introduction of a linear dependence of the phase and frequency of the external force on the dynamical variable in a non-autonomous linear oscillator significantly complicates the dynamics of such a simple system and leads to the emergence of a hierarchy of periodic and chaotic oscillations when the control parameters of the external influence are varied. The dynamics of such a system becomes close to a system with multi-well potential.

In the case of phase dependence on the dynamical variable, a hierarchy of chaotic attractors is observed resulting from cascades of period doubling bifurcations, while the lines of period doubling bifurcations are located at the doubled resonant frequency and subresonance frequencies. An increase in the amplitude of the external force leads to the expansion of the regions of existence of complex modes of oscillations; now these regions are not limited to twice the frequency of the external influence, as well as the appearance of new zones of periodic oscillations in the region of chaos. In this case, oscillation regimes appear in the dynamics of the system corresponding to the so-called dynamics of a nonlinear oscillator with a periodic potential.

In the case of the dependence of the frequency on the dynamical variable, a hierarchy of chaotic regimes is also observed, however, only the period doubling line remains in the vicinity of the doubled frequency of the external force. The system of bifurcation lines of period doubling at subresonance frequencies is destroyed with the formation of chaotic dynamics. However, an increase in the frequency of external force in this case also leads to the formation of a picture with a cascade of period doubling and the emergence of a hierarchy of chaotic regimes at the so-called superresonant frequencies.

The chaotic dynamics resulting from the control of the phase and frequency of a dynamic variable is characterized by broadband. The widest spectrum is observed in the case of a phase dependent on a dynamic variable, with the frequency of the external force near the resonance. For lower frequencies, the pronounced component of the base periodic signal is retained. For lower frequencies of external influence, the signal is filtered by the circuit and the spectrum has a limited band. In the case of a frequency dependence on a dynamical variable, the signal spectra are also broadband, however, the components of the basic limit cycle are pronounced.

**Acknowledgments.** This work was supported by the Russian Science Foundation, project No. 17-12-01008.
**Compliance with ethical standards.**
**Conflict of interest.** The authors declare that they have no conflict of interest.

### Reference
1. Best R. Phase-Lock Loops: Design, Simulation and Application, 6th ed. McGraw-Hill; 2007. p.1-489.
2. Shalfeev V D, Matrosov V V. Nonlinear dynamics of phase synchronization systems. Nizhny Novgorod: Publishing House of the Nizhny Novgorod State University; 2013. p.1-336. (in Russia)
3. Leonov G A, Kuznetsov N V, Yuldashev M V, Yuldashev R V. Hold-in, pull-in, and lock-in ranges of PLL circuits: rigorous mathematical definitions and limitations of classical theory. IEEE Transactions on Circuits and Systems I: Regular Papers 2015;62:2454-2464.
4. Ottesen J T. Modelling the dynamical baroreflex-feedback control. Mathematical and Computer Modelling 2000;31:167-173.
5. Hall J E. Guyton and Hall textbook of medical physiology e-Book. Elsevier Health Sciences; 2015. p.1-1147.
6. Ponomarenko V I, Prokhorov M D, Karavaev A S, Kiselev A R, Gridnev V I, Bezruchko B P. Synchronization of low-frequency oscillations in the cardiovascular system: Application to medical diagnostics and treatment. The European Physical Journal Special Topics 2013;222:2687–2696.
7. Kiselev A R, Karavaev A S, Gridnev V I, Prokhorov M D, Ponomarenko V I, Borovkova E I, Shvartz V A, Ishbulatov Y M, Posnenkova O M, Bezruchko B P. Method of estimation of synchronization strength between low-frequency oscillations in heart rate variability and photoplethysmographic waveform variability. Russian Open Medical Journal 2016;5:e0101-13.
8. Karavaev A S, Kiselev A R, Runnova A E, Zhuravlev M O, Borovkova E I, Prokhorov M D, Ponomarenko V I, Pchelintseva S V, Efremova T Yu, Koronovskii A A, Hramov A E. Synchronization of infra-slow oscillations of brain potentials with respiration. Chaos: An Interdisciplinary Journal of Nonlinear Science 2018;28:081102-5.
9. Verhulst F. Nonlinear differential equations and dynamical systems, 2nd edition. Springer Science & Business Media; 2006. p.1-311.
10. Kuznetsov A P, Kuznetsoc S P, Ryskin N M. Nelineynye kolebaniya, 2nd edition. Moscow: Fizmatlit; 2006. p.1-292. (in Russia)
11. Holmes P. A nonlinear oscillator with strange attractor. Phylos. Trans. 1979;292:419–448.

12. Humieres D D, Beasley M R, Huberman B A, Libhaber A. Chaotic states and rout to chaos in forced pendulum. Phys. Rev. A 1982;26:3484–3496.
13. Holmes P, Whitley D. On attracting set of Duffing`s equation. Physica D 1983;7:111–123.
14. Sato S, Sano M, Sawada Y. Universal scaling property in bifurcation structure of Duffing's and generalized Duffing's equation. Phys. Rev. A 1983;28:1654–1658.
15. Englisch V, Lauterborn W. Regular window structure of a double–well Duffing's oscillator. Phys. Rev. A 1991;44:916–924.
16. Toda M, Studies of a Non-Linear Lattice. Phys. Rep. 1975;18:1–123.
17. Kurz Th, Lauterborn W. Bifurcation structure of the Toda oscillator. Phys. Rev. A 1988;37:1029–1031.
18. Astakhov V V, Bezruchko B P, Seleznev E P. Investigations of the dynamics of an oscillatory circuit under harmonic excitation. Radiotekhnika i Elektronika 1987;32:2558-2566. (in Russia)
19. Bezruchko B P, Seleznev E P. Complex dynamics of a driven oscillator with a piecewise-linear characteristic. Technical Physics Letters 1994;20:800-801.
20. Bezruchko B P, Prokhorov M D, Seleznev E P. Multiparameter model of a dissipative nonlinear oscillator in the form of one-dimensional map. Chaos, Solitons & Fractals 1995;5:2095-2107.
21. Seleznev E P, Zakharevich A M. Structure of the control parameter space for a nonautonomous piecewise linear oscillator. Technical physics 2006;51:522-524.
22. Geist K, Lauterborn W. The Nonlinear Dynamics of the Damped and Driven Toda Chain: 1. Energy Bifurcation Diagrams. Phys. D 1988;31:103–116.
23. Geist K, Lauterborn W. The Nonlinear Dynamics of the Damped and Driven Toda Chain: 2. Fourier and Lyapunov Analysis of Tori. Phys. D 1990;41:1–25.
24. Geist K, Lauterborn W. The Nonlinear Dynamics of the Damped and Driven Toda Chain: 3. Classification of the Nonlinear Resonances and Local Bifurcations. Phys. D 1991;52:551–559.
25. Astakhov V V, Bezruchko B P, Kuznetsov S P, Seleznev E P. Onset of Quasiperiodic Motions in a System of Dissipatively Coupled Nonlinear Oscillators Driven by a Periodic External Force. Sov. Tech. Phys. Lett. 1988;14:16–18.
26. Kozlowski J, Parlitz U, Lauterborn W. Bifurcation Analysis of Two Coupled Periodically Driven Duffing Oscillators. Phys. Rev. E 1995;51:1861–1967.
27. Kuznetsov A P, Stankevich N V, Tyuryukina L V. The Death of Quasi-Periodic Regimes in a System of Dissipatively Coupled Van der Pol Oscillators under Pulsed Drive Action. Technical Physics Letters 2008;34:643-645.
28. Kuznetsov A P, Seleznev E P, Stankevich N V. Nonautonomous Dynamics of Coupled van der Pol Oscillators in the Regime of Amplitude Death. Commun. Nonlinear Sci. Numer. Simul. 2012;17:3740–3746.
29. Dvorak A, Kuzma P, Perlikowski P, Astakhov V, Kapitaniak T. Dynamics of three Toda oscillators with nonlinear unidirectional coupling. The European Physical Journal Special Topics 2013;222:2429-2439.
30. Stankevich N V, Dvorak A, Astakhov V, Jaros P, Kapitaniak M, Perlikowski P, Kapitaniak

T. Chaos and Hyperchaos in Coupled Antiphase Driven Toda Oscillators. Regular and Chaotic Dynamics 2018;23:120–126.
31. Josephson B D. Possible New Effects in Superconductive Tunnelling. Phys. Lett. 1962;1:251–253.
32. Salam F, Sastry S. Dynamics of the forced Josephson junction circuit: the regions of chaos. IEEE transactions on circuits and systems 1985;32: 784-796.
33. Cawthorne A B, Whan C B, Lobb C J. Complex dynamics of resistively and inductively shunted Josephson junction. J. Appl. Phys. 1998;84:1126–1132.
34. Dana S K, Chaotic Dynamics in Josephson junction. IEEE Transactions on circuits and systems-I: fundamental theory and applications 2001;48:990–996.
35. Sprott J C, Hoover W G. Harmonic oscillators with nonlinear damping. International Journal of Bifurcation and Chaos 2017;27:1730037-19.
36. Tang Y X, Khalaf A J M, Rajagopal K, Pham V T, Jafari S, Tian Y. A new nonlinear oscillator with infinite number of coexisting hidden and self-excited attractors. Chinese Physics B 2018;27:040502-6.
37. Jahanshahi H, Rajagopal K, Akgul A, Sari N N, Namazi H, Jafari S. Complete analysis and engineering applications of a megastable nonlinear oscillator. International Journal of Non-Linear Mechanics, 2018;107:126-136.
38. Vo T P, Shaverdi Y, Khalaf A J M, Alsaadi F E, Hayat T, Pham V T. A Giga-Stable Oscillator with Hidden and Self-Excited Attractors: A Megastable Oscillator Forced by His Twin. Entropy 2019;21:535-14.
39. Seleznev E P, Stankevich N V. Complex Dynamics of a Non-Autonomous Oscillator with a Controlled Phase of an External Force. Technical Physics Letters 2019;45:57-60.